\documentclass[onecolumn]{aastex63}
\usepackage{amssymb, amsmath}
\usepackage{color}
\usepackage{morefloats}
\usepackage{hyperref}

\begin{document}

\title{Newly Discovered Wolf-Rayet Stars in M31}
\email{Kathryn.Neugent@cfa.harvard.edu}

\author[0000-0002-5787-138X]{Kathryn F. Neugent}
\thanks{NASA Hubble Fellow}
\affiliation{Center for Astrophysics, Harvard \& Smithsonian, 60 Garden St., Cambridge, MA 02138, USA}

\author[0000-0001-6563-7828]{Philip Massey}
\affiliation{Lowell Observatory, 1400 W Mars Hill Road, Flagstaff, AZ 86001, USA}
\affiliation{Department of Astronomy and Planetary Science, Northern Arizona University, Flagstaff, AZ, 86011-6010, USA}

\begin{abstract}
The evolved massive star populations of the Local Group galaxies are generally thought to be well-understood. However, recent work suggested that the Wolf-Rayet (WR) content of M31 may have been underestimated. We therefore began a pilot project to search for new WRs in M31 and re-examine the completeness of our previous WR survey finished almost a decade prior. Our improved imaging data and spectroscopic follow-up confirmed 19 new WRs across three small fields in M31. These newly discovered WRs are generally fainter than the previously known sample due to slightly increased reddening as opposed to intrinsic faintness. From these findings, we estimate that there are another $\sim$60 WRs left to be discovered in M31; however, the overall ratio of WN-type (nitrogen-rich) to WC-type (carbon-rich) WRs remains unchanged with our latest additions to the M31 WR census. We are in the process of extending this pilot WR survey to include the rest of M31, and a more complete population will be detailed in our future work.
\end{abstract}
 
\section{Introduction}
Wolf-Rayet stars (WRs) are descended from the most massive of stars ($>30M_\odot$) and are characterized by strong stellar winds that produce broad emission lines in their spectra (e.g., \citealt{Cassinelli1975, Conti78}). WN-type, or nitrogen-rich WRs, show emission lines of helium and nitrogen, the products of CNO-cycle hydrogen burning. WRs that have undergone even more stripping, through enhanced stellar winds or binary interaction, will instead show the helium-burning products of carbon and oxygen (WC-type or WO-type). Once these massive stars end their stellar lives, they leave behind black holes. Around 30-40\% of WRs exist in short-period ($< 10$ day) binary systems \citep{Bartzakos2001, Foellmi2003, Schnurr2008, BinaryFrequency, Dsilva2022} and are the progenitors to the black hole mergers being detected as gravitational wave events \citep{LIGOPaperI}. However, while constraints have been placed on the binary fraction of the short-period WR systems, the overall influence of binarity on the evolution of WRs, and massive stars in general, is still an open question.

To understand the relative importance of binarity on the evolution of WRs, we first must identify a complete population of them in nearby environments, such as in the star-forming galaxies of the Local Group. Luckily for us, WRs are relatively straight-forward to identify observationally due to their characteristic strong and broad spectroscopic emission lines. Historically, our lack of knowledge about the WR populations within the Local Group was limited primarily due to observational resources rather than an inability to differentiate WRs from other stellar objects; see discussion in \citet{MJ98}, \citet{M33WRs}, and \citet{M31WRs}. Within the last few decades, we've led several targeted surveys to identify populations of WRs within the Local Group galaxies of M31, M33, and the Magellanic Clouds \citep{M31WRs,M33WRs,PaperI,PaperII,PaperIII,PaperIV} with reported completeness rates to within $\sim5\%$ or better. These surveys allow for exciting science to be done on galaxy-wide populations such as determining the physical and evolutionary properties of these stars \citep{Shenar2020, Shenar2016, Shenar2015}, understanding how WR populations relate to expected supernovae and black hole production rates (\citealt{Woosley2020}; Sarbadhicary et al., in prep), and investigating the impact of binary evolution by comparing the relative number of different types of massive stars across varying environments \citep{Trevor2018, RSGWR}.

In \citet{RSGWR}, we examined how the relative number of WRs and red supergiants (RSGs) should change as a function of metallicity and initial binary fraction within the Local Group. As is shown in Figure 11 of \citet{RSGWR}, we found generally good agreement between our observed WR to RSG ratio and model predictions from the Binary Population and Spectral Synthesis models (BPASS; \citealt{BPASS2}) for all galaxies except in M31, where the models suggest an initial binary fraction of 0\% (see additional discussion in Section 4 of \citealt{RSGWR}). This result would be surprising, but is currently not ruled out observationally. However, a more reasonable question might be: are we missing WRs in M31? This was not the first time we (and others) had pondered this question. \citet{SharaStar} found a reddened WR in M31 while searching for symbiotic stars and suggested that a ``modest number of reddened WR stars remain to be found in M31." At the time, we (incorrectly) assumed that the star discovered by \citet{SharaStar} had simply fallen in one of the chip gaps in our initial M31 study \citep{M31WRs} and concluded that this missing star was included in the 5\% completeness error. However, the discovery by \citet{SharaStar} and the conclusions found by \citet{RSGWR} motivated us to re-investigate the WR population of M31.

Here we present the results of this observational survey for additional WRs in M31. In Section 2 we identify the candidate WRs with interference filter imaging and image subtraction before obtaining their spectra. In Section 3 we give an overview of our new discoveries before discussing the broader implications of our survey's completeness in Section 4. Finally, we summarize our findings and offer concluding remarks in Section 5. Further information about our individual discoveries can be found in the Appendix.

\section{Identifying New WRs in M31}
Strong emission lines make WRs easily identifiable using narrow-band interference filter imaging. This method has been successfully used by us and others (e.g., \citealt{Wray1972, Moffat83, Armandroff85}) and is discussed extensively in \citet{M33WRs}. Here we summarize our procedure of using interference filters on the 4.3-m Lowell Discovery Telescope (LDT) to identify candidate WRs in M31 before spectroscopically confirming them using Binospec on the 6.5-m MMT.

\subsection{Candidate Identification with Interference Filter Imaging}
Following the methods used successfully in our initial M31 survey \citep{M31WRs}, we employed three interference filters optimized for the detection of WRs to be used on the LDT's Large Monolithic Imager (LMI). One filter is centered on the strongest emission line in WN-type WRs (He~{\sc ii} $\lambda4686$; {\it WN}), one is centered on the strongest line in WC-type WRs (C~{\sc iii/iv} $\lambda4650$; {\it WC}), and one is centered on neighboring continuum ($\lambda4750$; {\it CT}). The filter bandpasses are all 50\AA\ wide (full width at half maximum). Physically the filters are 120mm $\times$ 120mm to avoid vignetting the f/6 beam\footnote{The three filters were manufactured to our specifications by the Andover Corporation in September 2016, and have peak transmissions of $\sim80$\%.}. Objects that are brighter in the {\it WC} or {\it WN} filters compared to in the continuum are considered viable WR candidates worthy of potential spectroscopic follow-up. 

Our previous galaxy-wide interference imaging survey for WRs in M31 was conducted on the 4-m Mayall telescope at Kitt Peak. The Mosaic CCD had a much larger field of view compared to that of LMI on the LDT ($36\arcmin \times 36\arcmin$ vs.\ $12\farcm3 \times 12\farcm3$) and we were able to image the entire optical disk of M31 (2.2 deg$^2$) in 10 overlapping fields. Due to LMI's smaller field of view, we could not cover the same survey area in one observing season. Instead, we opted to image three fields with a higher S/N than what was obtained during the first survey, with the goal of both finding new WRs as well as quantifying the known shortcomings in our previous survey. 

There were two issues with the previous Mosaic survey done by \citet{M31WRs} that we hoped to improve upon. The first was due to the gaps within the 8-chip Mosaic CCD. While the images were dithered in an attempt to fill in the chip gaps, and the individual fields overlapped spatially, there were still regions missed. Additionally, poor cosmic-ray removal in the dithered regions hindered WR identification during the image subtraction process. As discussed in \citet{M31WRs}, we complemented our image subtraction techniques with photometry to identify stars that changed in magnitude between the on-band ({\it WC} and {\it WN}) and off-band ({\it CT}) images. The overall incompleteness due to gaps and cosmetics was estimated to be $\sim$5\% by \citet{M31WRs}; here we re-examine the issue in Section 4. 

The second issue was that the survey suffered from poor and variable seeing. Ideally, variable seeing (and thus point-spread-functions) should not pose a problem for image subtraction techniques due to the various cross-convolution methods employed. However, while completing a similar survey for WRs in the Magellanic Clouds \citep{PaperI,PaperII,PaperIII,PaperIV}, we found that the maximum difference in seeing between images had to be less than $0\farcs2$ or else the image subtraction routines lost effectiveness. Because this wasn't well understood during the original Mosaic M31 observations, and observing time wasn't in unlimited supply, the seeing fluctuations varied up to $0\farcs5$ leading to individual fields with poor image subtraction results. 

\begin{figure}
\center
\includegraphics[width=0.5\textwidth]{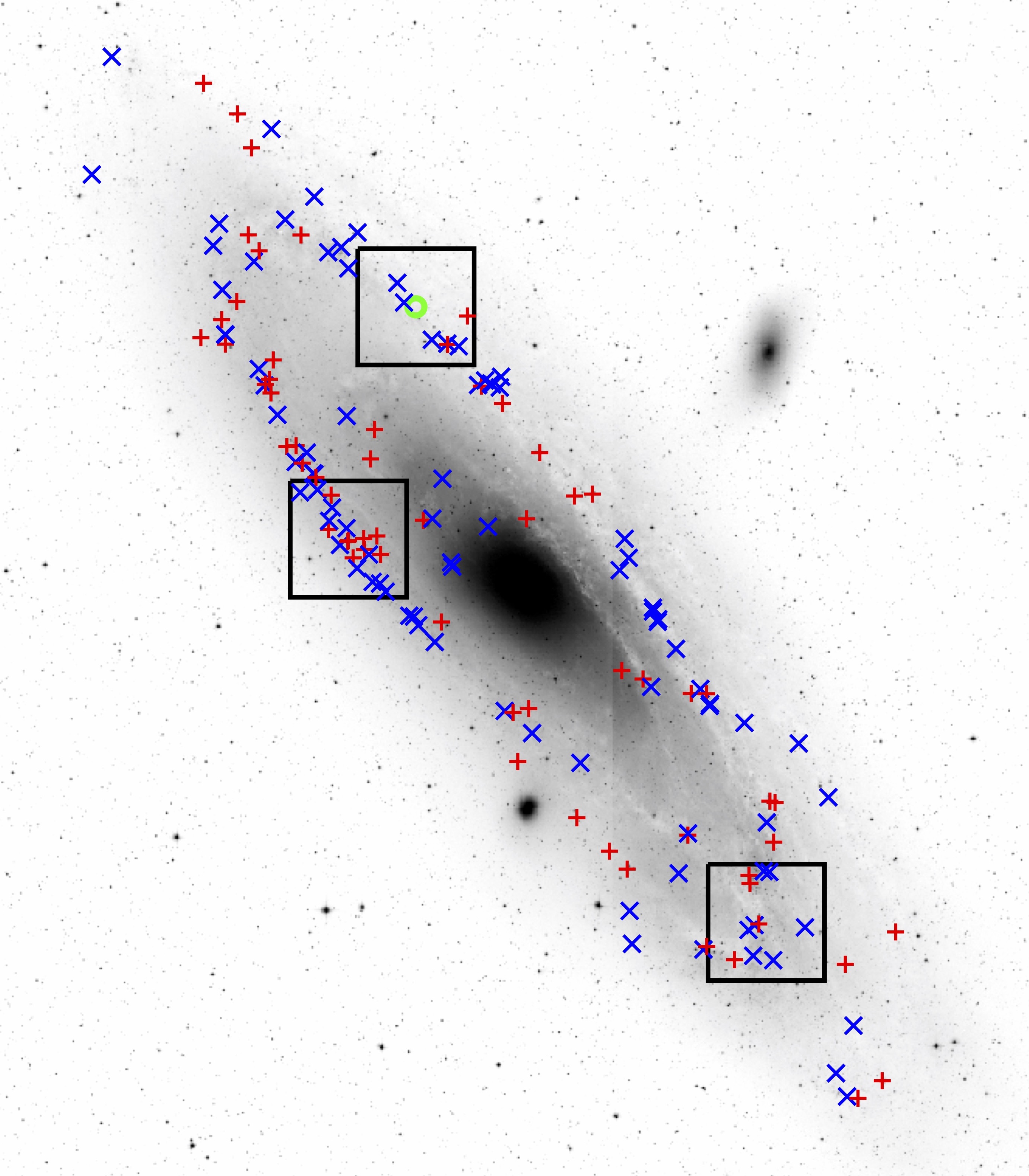}
\caption{\label{fig:fields} Locations of known WRs in M31 as well as new fields observed in this survey. The blue $\times$s represent WN stars while the red $+$s represent WC stars from \citet{M31WRs}. The green circle is the reddened WR discovered by \citet{SharaStar}. The three fields observed in this survey with LMI are denoted by black boxes.}
\end{figure}

Putting these two shortcomings together, we opted to re-image regions of M31 that were plagued both by chip gaps and unstable seeing during the original survey. We additionally focused on areas of M31 with dense OB populations and a large number of known WRs. We also made sure to include the reddened WR discovered by \citet{SharaStar} in an attempt to better understand if our new observational set-up would be able to detect it. The coordinates of our selected fields can be found in Table~\ref{tab:observations} and their locations within M31 relative to the known WRs are shown in Figure~\ref{fig:fields}.

\subsubsection{Observing and Image Reduction}
We observed three regions of M31 with LMI on the 4.3-m LDT during dark time in Fall 2021. Our overall goal was to obtain multiple 20 minute exposures in each of the {\it WN}, {\it WC}, and {\it CT} filters and then produce final, combined, images to identify candidate WRs using image subtraction and photometry. However, due to technical issues and variable seeing, we were not entirely successful. For all three fields, we managed to obtain at least one 20 minute image in the {\it WN} and {\it CT} filters with consistent seeing. In two fields we additionally obtained a shorter {\it WC} image as well as additional 20 minute exposures with the {\it WN} and {\it CT} filters. While the lack of a {\it WC} image for Field 1 was disappointing, this did not hinder our ability to identify WR candidates; it just made it more difficult to differentiate them from planetary nebulae (PNe) before spectroscopic follow-up, as discussed below. A summary of our fields, dates observed, exposure times, and seeing conditions can be found in Table~\ref{tab:observations}. 

\begin{deluxetable*}{l l l l l l l l l}
\tabletypesize{\scriptsize}
\tablecaption{\label{tab:observations} Observation Summary}
\tablewidth{0pt}
\tablehead{
\colhead{Name}
&\colhead{Coordinates}
&\colhead{Date Obs.\ (UT)}
&\colhead{Seeing}
&\colhead{Filter}
&\colhead{Exp.\ Time (s)}
}
\startdata
Field 1 & 00:43:46.40 +41:46:11.3 & 2021 Sep 8 & $0\farcs9$ & & &\\
& & & & {\it WN} & 2$\times$1200\\
& & & & {\it CT} & 2$\times$1200\\
& & & & {\it WC} & \nodata\\
Field 2 & 00:44:25.00 +41:21:00.0 & 2021 Sept 9 & $0\farcs8$ & & &\\
& & & & {\it WN} & 2$\times$1200\\
& & & & {\it CT} & 2$\times$1200\\
& & & & {\it WC} & 900\\
Field 3 & 00:40:25.00 +40:39:27.0 & 2021 Oct 4 & $0\farcs8$ & & &\\
& & & & {\it WN} & 1200\\
& & & & {\it CT} & 1200\\
& & & & {\it WC} & 900\\
\enddata
\end{deluxetable*}

The LMI detector is a 92.2-mm by 92.4-mm single array of 6144$\times$6160 e2v CCD231-C6 with a multilayer AR-coating\footnote{Although this detector series is now in use at many other observatories, ours was the first of its kind. Some of the developmental engineering costs were supported by the purchase price and covered by a generous grant from the National Science Foundation through AST-1005313.}. It is cooled to -120 C using a Stirling closed-cycle cooler. Further details can be found in the instrument manual\footnote{\url{http://www2.lowell.edu/users/massey/LMIdoc.pdf}}.  All images were taken with the detector binned by 2$\times$2, resulting in a scale of 0\farcs24 pixel$^{-1}$, and covering a field-of-view of 12\farcm3$\times12\farcm3$.  Although the detector has four excellent amplifiers, we chose to read our images out through a single one, simplifying the reduction process; in binned mode, the read-out time is only 24 seconds, small compared to our 20-minute exposure times. The read-noise is 6e, and the gain is 2.9 e/ADU. The detector is operated at voltages in which it is linear to a fraction of a percent up to 150,000 e- and beyond.

Each image contains a 32-column overscan region, which was used to remove frame-to-frame dependent bias levels. There remains a significant bias structure, which we removed by averaging ten bias frames, taken daily, although in practice the bias structure is stable on the time scale of months or even years. Multiple twilight flats were taken through each filter each evening with the telescope dithered between exposures to facilitate the removal of any stars bright enough to affect the flats. These exposures were averaged after scaling by their mode, and normalized by dividing by the median image value. Although the chip is otherwise excellent, there is a single bad column which we ignored.  All of these basic reductions took place using {\sc iraf}.  A ``pretty good" astrometric solution (0\farcs3) was added to each of the science images using the astrometry.net software \citep{2010AJ....139.1782L}.

\subsubsection{Candidate Identification}
To photometrically identify candidate WRs, we ran our own automated {\sc iraf} scripts to perform PSF-fitting photometry (based on {\sc daophot}; see \citealt{1987PASP...99..191S}). These scripts are adaptations for LMI of the programs used for the Local Group Galaxy Survey (LGGS) photometry described in \citet{2006AJ....131.2478M}. Stars whose photometry was statistically significantly brighter in the on-band filter compared to the off-band filter were identified as potential WRs candidates, using the same methodology described in \citet{M33WRs, M31WRs}. We emphasize that our photometry proved incidental to the discovery process, since many of the previously missed WRs were essentially invisible in the {\it CT} frame even in our new data, as described in the next section.

\begin{figure}
\center
\includegraphics[width=0.75\textwidth]{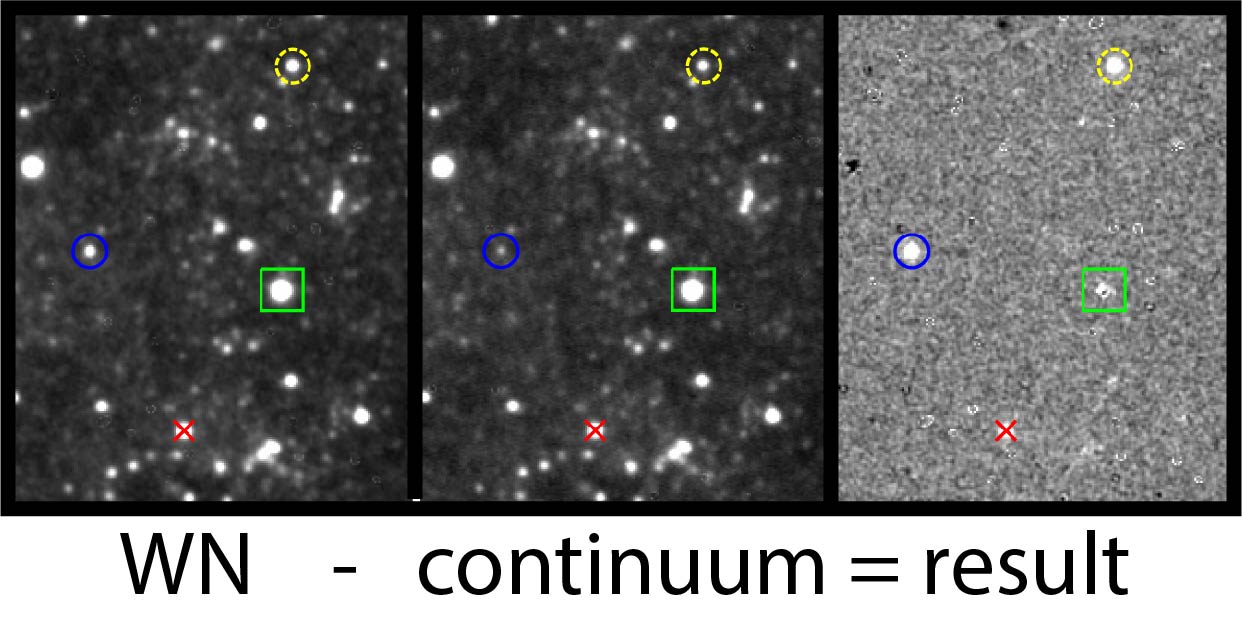}
\caption{\label{fig:subtract} Using image subtraction to identify WR candidates. We used {\sc hotpants} \citep{hotpants} to subtract the continuum image from the WN image to produce the final result. This particular region of the sky shows a previously known WR (yellow dashed circle), a saturated star that leads to a poor image subtraction result (green box), a M-type star that shows up due to TiO absorption in the continuum filter (red $\times$), and a new WR candidate that was missed in our previous survey due to its faint magnitude (blue solid circle).}
\end{figure}

For image subtraction, we used the High Order Transform of PSF And Template Subtraction ({\sc hotpants}) code written by \citet{hotpants}. This allowed us to ``subtract" the {\it CT} image from both the {\it WC} and {\it WN} images. Since WRs will have more flux in either the {\it WC} or {\it WN} images compared to in the {\it CT}, they will appear prominently in the subtracted image, as is shown in Figure~\ref{fig:subtract}. However, as is also shown in Figure~\ref{fig:subtract}, they are not the only stars that will show up in the subtracted images. Such contaminants are discussed below.

The three main non-WR contaminants that appear after image subtraction are saturated stars, red stars, and PNe. Saturated stars are relatively easy to discount as WR candidates because they are visually bright in all filters and the resulting image-subtracted ``star" often has a non-circular shape (see star in the green box in Figure~\ref{fig:subtract} for example). Red stars (generally foreground M dwarfs) appear as potential candidates because there is a TiO absorption line in the {\it CT} filter. This makes them appear brighter in the {\it WC} or {\it WN} filter compared to in the {\it CT}. While red stars were a large source of contamination in our first survey in M31 and M33 \citep{M31WRs, M33WRs}, this time we cross-matched our candidates with the LGGS to identify stars with $B-V < 0.7$, as that corresponds to the color of a G-type dwarf and should remove the majority of any stars with TiO bands \citep{AAQ}. Finally, PNe additionally appear as potential WR candidates because they show He~{\sc ii} $\lambda4686$ nebular emission if the radiation field is hard enough. The majority of PNe in M31 have been cataloged (e.g. \citealt{Ciardullo89, Merrett2006, Bhattacharya2019}) and thus it was possible to cross-match our candidates with known PNe. Additionally, the fainter PNe tend to only show up through the {\it WN} filter (due to the nebular emission) and show nothing on the {\it CT} or {\it WC} image. However, since we were attempting to find reddened and faint WRs, we retained any such candidates until we could cross-match them with the PNe catalogs. In total, we removed 17 known PNe from our candidate list. 

Using the subtracted images combined with the photometry, we created a list of potential WR candidates. After removing the various contaminates, we cross-matched our list with known WRs from \citet{M31WRs}. We successfully identified all of the known WRs in our three fields which included 11 known WRs in Field 1, 19 in Field 2, and 8 in Field 3. Additionally, the WR discovered by \citet{SharaStar} showed up prominently on the {\it WN} field, but was too faint for detection in the {\it CT} image. After removing the known WRs, we were left with 30 new WR candidates. These 30 candidates were then divided into high, medium, and low priorities based on their likelihood of being a WR. There were 11 high priority candidates that were detectable in both the {\it WN} and {\it WC} images (if available), as well as a weaker detection in the {\it CT}, and were thus likely not PNe. There were an additional 15 medium priority candidates that had a strong detection in the {\it WN}, a weak detection in the {\it WC} (if available), but no detection in the {\it CT}. This led us to conclude that they were likely either newly found PNe or reddened WRs of WN-type such as the one discovered by \citet{SharaStar} (due to the weak detection in the {\it WC}). Finally, there were 4 remaining candidates that had very small photometric differences between the on and off band filters, but could potentially be Of-type stars and were still worth spectroscopically confirming. 

After identifying our list of candidates with the LMI images, we needed better astrometry for the spectroscopic confirmation than the 0\farcs3 uncertainty that sufficed for the cross-identifications. Fortunately, LMI is often used for precise astrometry when determining the positions of asteroids and Kuiper Belt object for recovery or orbital stability studies (e.g., \citealt{2022NatCo..13..447S}), or for orbital refinements in order to predict occultation tracks accurately (e.g.,\citealt{2021AJ....161..210L}) -- both far more demanding tasks than ours. To facilitate such studies, care was taken in the design of the image corrector to minimize distortions that would affect the astrometry \citep{2012SPIE.8444E..51B,2014SPIE.9145E..2CD}.  The final astrometry of the WR candidates were determined using IDL routines written by Marc Buie\footnote{\url{https://www.boulder.swri.edu/~buie/idl/}}, with Gaia stars used as a reference \citep{GaiaDR2,GaiaEDR3}. Transformations for each WN image were computed using $\sim$4500 Gaia stars using an 11th order equation for the standard coordinates $\eta$ and $\xi$ with residual scatters of 0\farcs05.

\subsection{Spectroscopic Confirmation}
To spectroscopically confirm the candidate WRs in M31, we used Binospec on the 6.5-m MMT located on Mt.\ Hopkins in southern Arizona \citep{Binospec}. Its two $8' \times 15'$ fields of view were a good match to our three LMI fields and we were able to observe all 30 of our candidate WRs in three Binospec fields. They were observed during dark time on UT 2022 Oct 26 (Field 1), 2022 Oct 31 (first half of Field 2), 2022 Dec 14 (second half of Field 2), 2022 Dec 25 (first half of Field 3), and 2022 Dec 26 (second half of Field 3). The seeing ranged from $0\farcs9 - 2\farcs0$ with the majority of the observations taken with a seeing of $<1\farcs2$, which was a reasonable match to our slit width of $1\farcs0$. Fields 1 and 2 were observed with clear sky conditions while Field 3 was observed with clouds. Exposure times were 3 hours long (broken up into $9\times20$ minute exposures) to achieve a S/N of 30 per 3.3\AA\ resolution element at $\sim4500$\AA\, which we determined was adequate for classifying WRs in M31 in \citet{M31WRs}. 

The optimal wavelength coverage to classify WRs is $4000-5900$\AA\ which includes N\,{\sc iv} $\lambda 4058$, N\,{\sc v} $\lambda\lambda 4603,19$, N\,{\sc iii} $\lambda \lambda 4634,42$ and He\,{\sc ii} $\lambda 4686$ for WN-types, and C\,{\sc iii/iv} $\lambda 4650$, O\,{\sc v} $\lambda 5592$, C\,{\sc iii} $\lambda 5696$, and C\,{\sc iv} $\lambda 5806$ for WC-types. We opted to observe our candidates using the 270 l/mm grating which gave us sufficient resolution, adequate sensitivity in the blue, and a $\sim5400$\AA\ wavelength range. Depending upon where a slit was located in the field, the wavelength shift could be as much as $\pm$ 1300\AA, so we used a minimum central wavelength of 5501\AA\ to achieve as much coverage in the blue wavelength regime as possible. As a result, while the starting and ending wavelengths vary for each of the spectra, the minimum wavelength coverage ranges from $4020-7360$\AA. The spectra were reduced on a night-by-night basis using the Binospec pipeline \citep{Kansky2019}. Due to variable quality, we chose not to combine spectra from multiple nights. Instead, we present the best versions in FITS format as ``data behind figures" in the online Journal with the final exposure times for each spectrum available in the headers.

In an effort to fill the masks with targets, we additionally re-observed 22 known WRs in M31. A list of the known WRs we re-observed are in Table~\ref{tab:knownWRs} with updated spectral types, as applicable. Additionally, these spectra are also available in FITS format as ``data behind figures" in the online Journal.

\begin{deluxetable*}{l l l l}
\tablecaption{\label{tab:knownWRs} Re-Observed Known WRs}
\tablewidth{0pt}
\tablehead{
\colhead{ID}
&\colhead{Old Sp. Type\tablenotemark{a}}
&\colhead{New Sp. Type}
&\colhead{Comments}
}
\startdata
J004020.44+404807.7 & WC6  & WC6 & \\ 
J004022.43+405234.6 & WC4  & WC4 & \\
J004023.02+404454.1 & WN4  & WN6 & N\,{\sc iii}$\sim$N\,{\sc iv}; N\,{\sc v} present but weak\\ 
J004026.23+404459.6 & WN6+abs & WN6+abs & \\
J004031.67+403909.0 & WN6  & WN5-6 & \\
J004034.17+404340.4 & WC7+fgd & WC7+fgd & \\ 
J004034.69+404432.9 & WC4 & WC5? & Weak C\,{\sc iii}? \\
J004109.46+404907.8 & WN6 & WN6 & \\
J004302.05+413746.7 & WN9: & WN8 & No N\,{\sc ii} \\
J004316.44+414512.4 & WC6 & WC7 & \\
J004337.10+414237.1 & Ofpe/WN9 & Ofpe/WN9 & \\
J004349.72+411243.4 & WN6 & WN6 & \\
J004353.34+414638.9 & WN7 & WN6 & \\
J004357.31+414846.2 & WN8 & WN8 & \\
J004403.39+411518.8 & WN6 & WN4.5-5 & \\
J004408.58+412121.2 & WC6 & WC6 & \\
J004410.17+413253.1 & WC6 & WC6 & \\
J004410.91+411623.2 & WN4 & WN4 & \\
J004412.44+412941.7 & WC6 & WC6 & \\
J004422.24+411858.4 & WC7-8 & WC7 & \\
J004434.57+412424.4 & WN3 & WN4 & N\,{\sc iv}$\sim$N\,{\sc v}; N\,{\sc iii} absent\\
J004436.22+412257.3 & WN5 & WN6 & \\
\enddata
\tablenotetext{a}{Classifications from \citealt{M31WRs} and references therein.}
\end{deluxetable*}

\section{The Newly Discovered Wolf-Rayets}
\begin{figure}
\center
\includegraphics[width=1\textwidth]{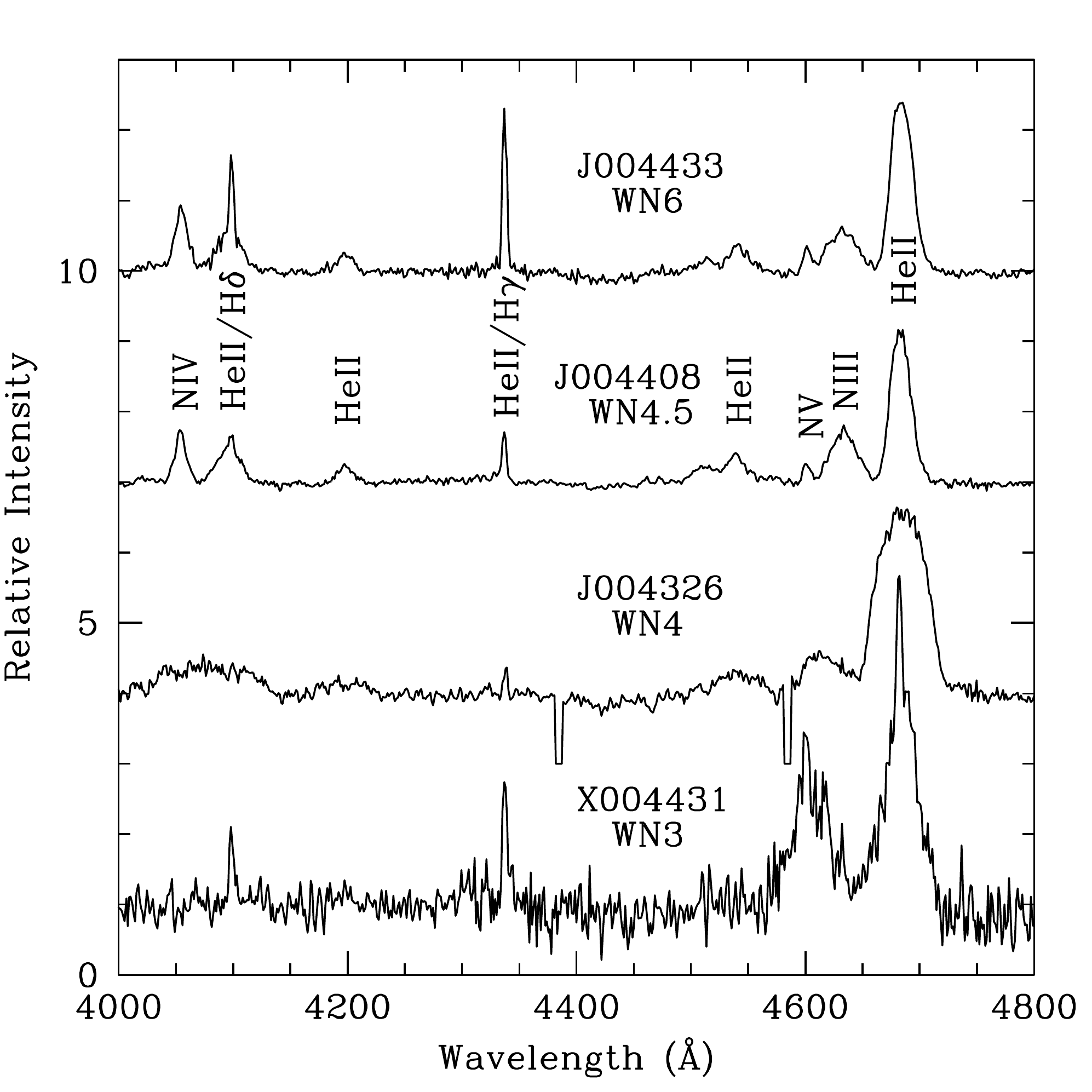}
\caption{\label{fig:WNspec} Spectra of representative WNs. We show four example spectra of newly found WN stars in M31. The spectra have been normalized, and only a portion of our wavelength
coverage is shown for clarity. Each spectrum is offset from the previous by a relative intensity of 3.0.
Lines identified include N\,{\sc iv} $\lambda 4058$, the He\,{\sc ii}/H$\delta$ blend (contaminated by nebular emission in X004431.39+412114.0 and J004433.91+412501.2), He\,{\sc ii} $\lambda 4200$, the He\,{\sc ii}/H$\gamma$ blend (with similar nebular contamination), He\,{\sc ii} $\lambda 4542$, and the N\,{\sc v} $\lambda \lambda 4603,19$ and N\,{\sc iii} $\lambda 4634,42$ doublets, as well
as He\,{\sc ii} $\lambda 4686$. Note that the unnormalized version of these, and all of our spectra, are being made available in FITS format as part of this publication as ``data behind figures".}
\end{figure}

\begin{figure}
\center
\includegraphics[width=1\textwidth]{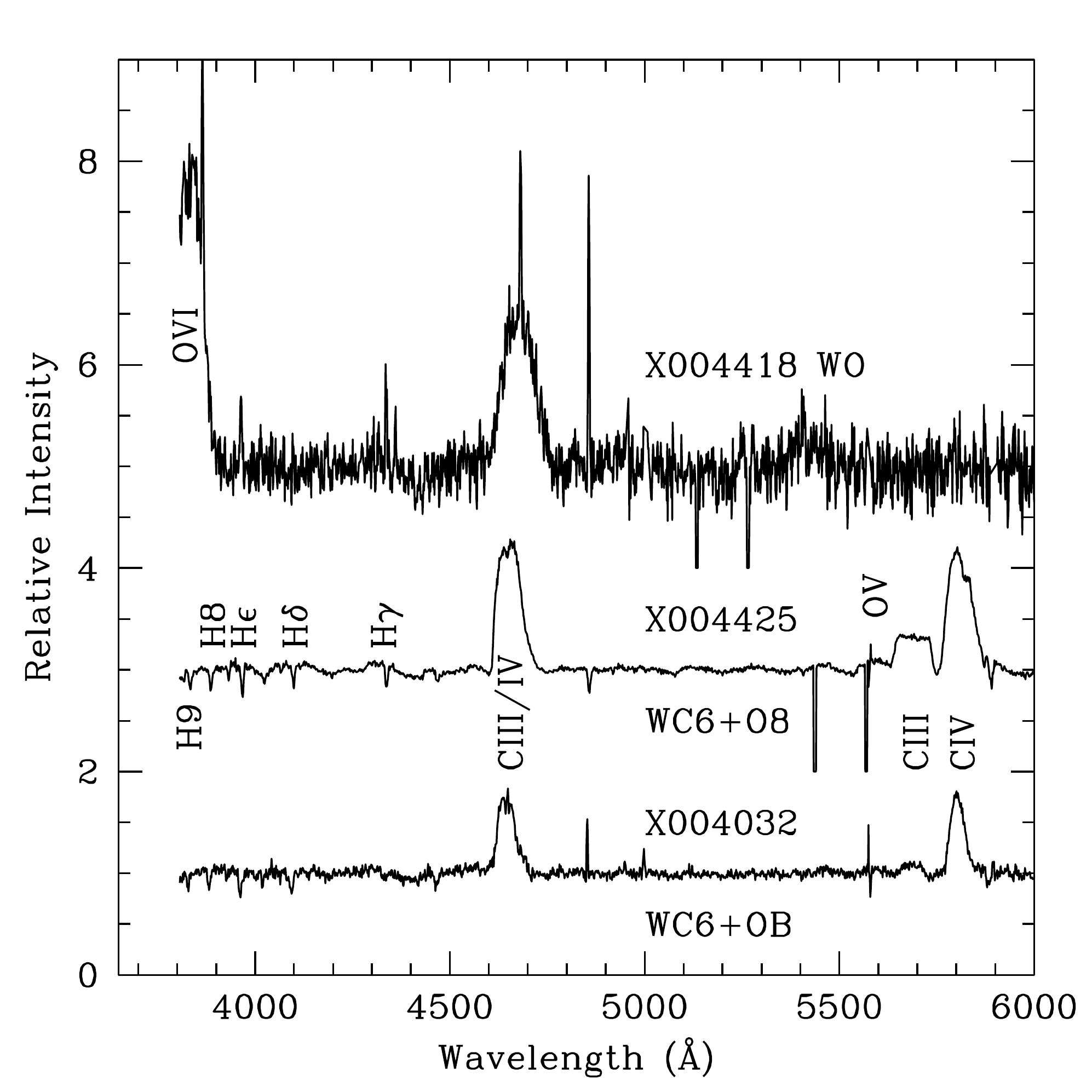}
\caption{\label{fig:WCspec} Spectra of representative WC/WOs.  We show three example spectra of newly found WC/WO stars in M31. The spectra have been normalized, and only a portion of our wavelength coverage is shown for clarity.  Each spectrum is offset from the previous by a relative intensity of 2.0. Lines identified include the O\,{\sc vi} doublet $\lambda \lambda 3811,34$ (X004418.10+411850.8),  the upper Balmer lines (visible in X004032.57+403901.3 and X004425.09+412046.4), the  C\,{\sc iii/iv} $\lambda 4650$ feature, and the classification lines O\,{\sc v} $\lambda 5592$, C\,{\sc iii} $\lambda 5696$, and C\,{\sc iv} $\lambda 5806$. Note that the unnormalized version of these, and all of our spectra, are being made available in FITS format as part of this publication as ``data behind figures".}
\end{figure}

Of the 30 WR candidates we identified, 19 are newly discovered WRs in M31. Of the remaining eleven candidates, one is an Of-type star with He\,{\sc ii} $\lambda4686$ emission (discussed more below), six are H\,{\sc ii} regions with nebular He\,{\sc ii} $\lambda4686$, three are cool stars with $B-V$ colors slightly below our red color-cutoff, and one is an A-type star that appeared as a candidate likely due to its brightness and resulting poor image subtraction. These stars are listed in Table~\ref{tab:specObs} and are discussed individually in the Appendix. Additionally, spectra of a few representative WNs are shown in Figure~\ref{fig:WNspec} and WCs in Figure~\ref{fig:WCspec}.

The WR classification criteria are nicely enumerated in Table 2 of \citet{2001NewAR..45..135V}. These criteria are primarily qualitative since the various complexities surrounding WR winds (mass-loss rates, density and velocity profiles, etc.) dictate that two WRs with the same subtype may still have line fluxes that differ by large amounts. Other line ratios may also differ significantly. Therefore, while there are no WR spectral ``standards", we can still strive for consistency across classifications. The digital versions of similarly classified WRs in M31 and M33 can be found in a tar file published as part of \citet{M31WRs}. Furthermore, digital optical spectra of most of the single Galactic WN stars described by \citet{Hamann95} can be found here\footnote{\url{https://www.astro.physik.uni-potsdam.de/~www/research/abstracts/wn-atlas_abstract.html}}. Finally, additional WN and WC stars from \citet{1988AJ.....96.1076T} can be downloaded from Vizier.

The spectral subtypes of WN stars are based upon the relative strengths of N\,{\sc iii}~$\lambda\lambda4634,42$, N\,{\sc iv}~$\lambda4058$, and N\,{\sc v}~$\lambda\lambda4603,19$, with low types (such as WN3) assigned to stars where N\,{\sc v} dominates, and later types (such as WN9) where N\,{\sc iii} dominates, with the strength of N\,{\sc iii} relative to He\,{\sc ii}~$\lambda4686$ serving as a secondary criterion to distinguish the WN7 from the WN8 class. The evolutionary connection between these classes is not clear owing to the disconnect between what is happening in the interior of these stars and the ionization balance in the outflowing wind where the emission lines are produced. Generally, the late-type WNs show the presence of hydrogen, while earlier types do not. (Hydrogen readily reveals itself as the even-N Pickering lines of He\,{\sc ii} are coincident with the Balmer lines. Both the He\,{\sc ii} and Balmer lines will show a decrease in strength going to shorter wavelengths, but if there is an odd/even pattern to the progression, with the even-N Pickering lines being stronger, that shows that hydrogen is still present; see \citealt{1983ApJ...268..228C}.) 

Similarly the WC subtypes are determined from the relative strengths of C\,{\sc iv}~$\lambda\lambda5801,12$, C\,{\sc iii}~$\lambda5696$, and O\,{\sc v}~$\lambda\lambda5592$. Again, the evolutionary connection between these ionization-based classifications is a topic of on-going work; what is clear is that the WC stars are more chemically evolved than the WN stars. If O\,{\sc vi}~$\lambda\lambda3811,34$ is present and {\it strong}, in some poorly defined sense, an early WC star is classified as a WO star. Recent work has shown that these WO stars are more evolved with most of their helium converted to carbon and oxygen \citep{2022ApJ...931..157A}. Finally, there are some WN stars that show abnormally strong C\,{\sc iv}~$\lambda\lambda5801,12$. These are often referred to as ``transition stars,” with the suggestion that they are WNs caught in the act of evolving to WCs, but the evidence here is also tenuous; see discussion in \citet{2021MNRAS.503.2726H}.

In terms of names, early studies of the M31 WR content used a hodge-podge of designations (see summary in Table 10 of \citealt{MJ98}). All of the newly found WRs by \citet{M31WRs}, as well as the previously known ones, were in the LGGS, and they adopted these coordinate-based designations. The exceptions were two WC6 stars,
which were not in the LGGS. For these, \citet{M31WRs} used LGGS-like designations but with an ``X" instead of a ``J" to avoid confusion.  SIMBAD lists these as ``[NMG2012] X J004256.05+413543.7" and ``[NMG2012] X J004308.25+413736.3", defeating our effort at simplification. Of our 20 newly found interesting stars, only 8 appear in the LGGS and thus we continue the ``X" nomenclature here.

In Table~\ref{tab:specObs} we include accurate coordinates and, when available, the LGGS designations. Note that there is a small offset between the given coordinates and the LGGS names. The LGGS names were tied to the USNO-B coordinates, while the coordinates of our newly found stars are on the ICRS, via Gaia, as described above. We also give the approximate equivalent widths of He\,{\sc ii} $\lambda4686$ (WN stars) or C\,{\sc iii/iv} $\lambda4650$ (WC stars), along with their full widths at half maximum (FWHM) line widths.  For comparison, the H\,{\sc ii} regions in our sample have FWHMs of 4.5\AA\ at similar wavelengths, consistent with the expected resolving power of $R\sim 1340$.

\begin{deluxetable*}{l l l l l l}
\tabletypesize{\scriptsize}
\tablecaption{\label{tab:specObs} Spectroscopic Results of New WR Candidates}
\tablewidth{0pt}
\tablehead{
\colhead{Designation}
&\colhead{RA}
&\colhead{Dec}
&\colhead{EW\tablenotemark{a}}
&\colhead{FWHM\tablenotemark{a}}
&\colhead{Classification}
}
\startdata
J004019.66+404232.5 & 00:40:19.630 & +40:42:32.40 & -100 & 28 & WN3\\
J004031.21+404128.1 & 00:40:31.190 & +40:41:28.04 & -130 & 18 & WN7\\
X004032.57+403901.3 & 00:40:32.566 & +40:39:01.30 &  -40 & 50 & WC6+OB\\
J004039.59+404449.5 & 00:40:39.574 & +40:44:49.42 &  -80 & 27 & WN3 \\ 
J004042.44+404505.3 & 00:40:42.423 & +40:45:05.28 & -100 & 23 & WN4.5 \\
X004045.57+404526.4 & 00:40:45.567 & +40:45:26.42 & \nodata & \nodata & \nodata \\
X004054.05+403708.3 & 00:40:54.047 & +40:37:08.29 & -900 & 50 & WC6 \\
X004318.88+414711.1 & 00:43:18.876 & +41:47:11.08 & \nodata & 4.1 & H\,{\sc ii} region\\ 
X004320.88+414107.0 & 00:43:20.884 & +41:41:07.02 & \nodata & \nodata & K7-M0\\
J004326.06+414260.0 & 00:43:26.022 & +41:42:59.85 & -130 & 46 & WN4 \\
X004332.04+414817.2 & 00:43:32.036 & +41:48:17.16 & \nodata & 4.6 & H\,{\sc ii} region\\ 
X004338.95+414327.0 & 00:43:38.947 & +41:43:26.98 & \nodata & 26: & WN \\
X004341.15+414413.6 & 00:43:41.148 & +41:44:13.65 & -25: & 20: &  WN8/C \\
X004353.03+412141.0 & 00:43:53.032 & +41:21:41.03 & -75  & 17  & WC \\
X004359.44+414823.9 & 00:43:59.439 & +41:48:23.87 & -30 & 25 & WN6 \\
J004404.10+411710.5 & 00:44:04.101 & +41:17:10.37 & \nodata & \nodata & H\,{\sc ii} region \\
X004406.41+412020.8 & 00:44:06.409 & +41:20:20.78 & -150 & 25 & WN4.5 \\
J004408.13+412100.6 & 00:44:08.113 & +41:21:00.56 &  -48 & 20 & WN4.5 \\
J004413.56+412004.7 & 00:44:13.555 & +41:20:04.46 &  -49 & 30 & WN4.5 \\
X004414.71+414033.3 & 00:44:14.708 & +41:40:33.32 & \nodata & 4.0 & Symbiotic? \\
X004418.10+411850.8 & 00:44:18.101 & +41:18:50.76 & -190 & 90 & WO \\
X004421.30+411807.2 & 00:44:21.301 & +41:18:07.16 &  -4 & 20 &  O6~If \\
X004423.98+412255.6 & 00:44:23.981 & +41:22:55.57 &  \nodata & \nodata & M2~I \\ 
X004425.09+412046.4 & 00:44:25.085 & +41:20:46.44 &  -95 & 65 & WC6+O8\\
X004426.99+411928.0 & 00:44:26.987 & +41:19:28.03 & \nodata & \nodata & A-type?\\
X004431.39+412114.0 & 00:44:31.393 & +41:21:14.04 &  -175 & 35 & WN3 \\ 
X004432.06+411940.5 & 00:44:32.059 & +41:19:40.46 &  \nodata & \nodata & H\,{\sc ii} region.\\
J004433.91+412501.2 & 00:44:33.903 & +41:25:01.13 &  -65 & 22 & WN6\\
X004438.04+412518.7 & 00:44:38.038 & +41:25:18.73 &  -45 & 16 & WN6 \\
X004440.49+412052.1 & 00:44:40.486 & +41:20:52.11 &  \nodata & 4.3 & H\,{\sc ii} region \\
\enddata
\tablenotetext{a}{Equivalent Width (EW) and Full Width at Half Maximum (FWHM), both in units of \AA, for He\,{\sc ii} $\lambda4686$ for WN stars, and C\,{\sc iii/iv} $\lambda4650$ for WC stars.}
\end{deluxetable*}

\section{Discussion}
The confirmation of 19 newly discovered M31 WRs from our three LMI fields raises a number of important questions. Why were these new WRs missed as part of our previous, galaxy-wide survey with the Mosaic camera \citep{M31WRs}? Were they missed because they fell on the inter-chip gaps of the Mosaic camera, or are they fainter and required a higher sensitivity to be detected? If they're fainter, is this because the new WRs are {\it intrinsically} fainter, or are they simply more heavily reddened? And finally, is this new survey deep enough, or if we increased our survey sensitivity, would we simply keep finding more and more WRs? We must answer these questions quantitatively before estimating the total number of missing WRs in M31. Finally, as is discussed at the end of this section, the discovery of additional WRs in M31 raises the question of whether or not our complementary Mosaic survey in M33 \citep{M33WRs} is also incomplete.

\subsection{Why were the previous WRs missed?}
We begin by addressing the first of these questions: why were these 19 M31 WRs missed by \citet{M31WRs}? The Kitt Peak Mosaic Camera used in that study consisted of eight 2K$\times$4K CCDs, with gaps between the chips. Each field was surveyed by obtaining nine exposures, three through each of the three narrowband filters ({\it WN}, {\it WC}, and {\it CT}) with the telescope offset (dithered) between exposures to help fill in the gaps. For each exposure, the gaps (and bad columns) affected, on average, $\sim$5\% of the 36\arcmin$\times$36\arcmin\ area covered by the exposure. The dithering process resulted in decreased sensitivity in three times this area when the data were combined. Thus the affected area is $\sim$15\%. This is ameliorated by the large overlap between adjacent fields. When we do a careful accounting, we expect about 9\% of the survey area was affected. However, we likely did not lose {\it all} WRs within this 9\%, as the gaps are {\it partially} filled in thanks to the dithering. Additionally, WRs with large differences between the on- and off-band magnitudes were detectable photometrically even in these dithered regions. \citet{M31WRs} estimate an incompleteness of about 5\% due the gaps and other cosmetics, and our re-evaluation here confirms that this is a reasonable estimate, with the incompleteness being preferentially biased towards the fainter WRs.

When we examine the location of the newly confirmed WRs on our old Mosaic data, we find that three fell into gaps, namely, J004031.21+404128.1, J004326.06+414260.0, and J004413.56+412004.7. We note that 39 of the previously known WRs fall on the LMI fields; thus having missed 3 stars due to gaps is in reasonable accord with our expectations of 5-10\%. An additional new WR, X004425.09+412046.4, was missed due to crowding. The other 15 new WRs were missed as part of the previous survey because they were either very faint or invisible on the Mosaic images. Thus, the vast majority ($\sim$80\%) of the newly found WR stars were not found in our earlier survey to a lack of sensitivity. However, this raises a new question: are these new WRs intrinsially fainter than the previously known population of M31 WRs, or is their faintness caused by increased reddening?

\begin{figure}
\center
\includegraphics[width=0.6\textwidth]{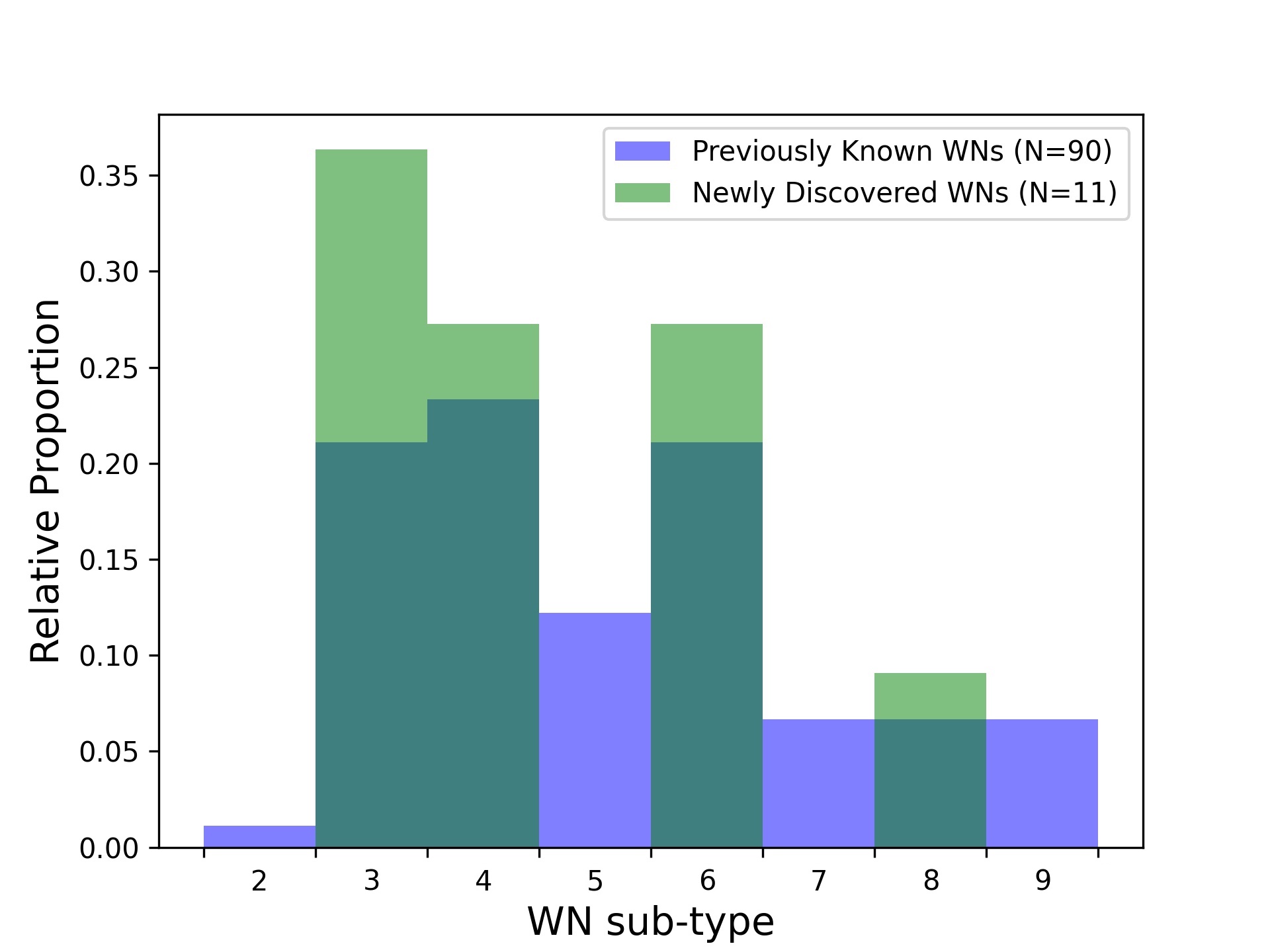}
\caption{\label{fig:WNsubtypes} The relative proportion of WN subtypes compared to both the previously known and newly discovered sample of M31 WNs. While the total number of stars in each group differs, their distributions are fairly similar. This suggests that the newly discovered WRs in M31 are not predominantly part of an intrinsically fainter group of WRs.}
\end{figure}

To investigate whether these new WRs are intrinsically fainter than the previously known M31 WRs, we can compare their subtype distributions and infered absolute magnitudes. In general, WN-type WRs are more difficult to detect than WC-type WRs due to weaker emission line fluxes (see e.g., \citealt{Conti89} and \citealt{MJ98}), so we focus on the WN-type WRs for this comparison. Figure~\ref{fig:WNsubtypes} shows the relative proportion of WN subtypes for both the previously known WNs in M31 ($N = 90$) as well as for the newly discovered sample of faint WNs ($N = 11$ due to removing the WCs and brighter WNs missed due to being in the Mosaic gaps). For reference, the absolute magnitudes of these stars range from $M_V = -3$ for the early-type WNs up to $M_V = -6/-7$ for the late-type WNs \citep{Crowther07}. While the total number of stars differs by nearly a factor of 9 between the two groups, the overall subtype distribution of the newly discovered sample is similar to that of the known sample. Thus, these new stars are not intrinsically fainter as a group. 

We can next examine the overall reddening of the newly discovered faint WRs. Due to their magnitudes, the majority of them are not in the LGGS \citep{2006AJ....131.2478M} and thus we do not know their $B-V$ or $E(B-V)$ values. However, 61 of the previously known M31 WRs and 10 of the new WRs are within the Panchromatic Hubble Andromeda Treasury (PHAT) footprint \citep{PHAT}. Of these 10 new WRs, 7 are part of the fainter sample that was missed with the previous Mosaic survey. \citet{Schlafly2011} computed extinction values for the various ACS and WFPC3 filters. We therefore can compare the {\it difference} in the observed colors (e.g., F275W-F336W) between the previously known M31 WRs stars and the newly found faint sample, and compute the corresponding extra amount of reddening between the two samples. From F275W-F336W, F336W-F475W, and F475W-F814W we find very consistently that the difference in colors between the two sets corresponds to an average differences of 0.25 in $E(B-V)$, with the newly found faint sample being more highly reddened. This corresponds to roughly an additional 0.95~mag in {\it WN}. Therefore, these new WRs were primarily missed in our previous survey due to increased reddening causing them to be fainter than the Mosaic detection limit. 

We next briefly discuss the source of this increased reddening. While WRs have high mass-loss rates ($10^{-6}$ to $10^{-4} M_\odot$ yr$^{-1}$; see \citealt{Sander20} and references therein), this doesn't necessarily lead to high {\it local} extinction values. Overall, very few dust-enshrowded WRs have been found (see \citealt{Barniske2008} for a few exceptions and an overview), though late-type WCs have been known to create carbon-rich dust in high metallicity environments \citep{Crowther06, Williams2021}. Additionally, binary WRs with highly ellipitical orbits have been observed episodically creating dust during periastron passages (see, for example, the recent images of WR140 from JWST \citealt{Lau2022}). However, WRs that are heavily extincted due to their own dust seem to be the exception rather than the rule. Thus, the reddened WRs we've observed are likely located in regions of globally higher reddening due to M31's inclined orientation relative to our viewing angle. 

How common are these regions of higher reddening in M31? \citet{1986AJ.....92.1303M} investigated the massive star content of multiple OB associations in M31 using both photometry and spectroscopy. They found reddening values varying from $E(B-V)=0.08$ (OB 102), $0.12$ (NGC 206) up to $0.24$ (OB 48 and the OB 8, 9, and 10 complex). They inferred that the $A_V$ extinction of these early-type stars ranged from 0.25-0.75~mag. \citet{M3133RSGs} argues that that these low numbers probably do not represent the typical extinction of the disk of M31, but rather adopts the higher $A_V\sim1.0$ value found from PHAT data by \citet{PHATdust}. \citet{Wang2022} finds the same typical $A_V\sim1.0$ value. Both \citet{PHATdust} and \citet{Wang2022} argue that there are regions with much higher extinctions, but that these represent a tiny fraction ($<$1\%) of the area of M31. As we discuss below, we hope to better quantify the reddening of these stars by obtaining new {\it B},{\it V}, and {\it R} images with LMI during our upcoming observing runs. Additionally, once we have more coverage of WRs within the PHAT survey area, we can compare our results to that of M31 dust extinction maps such as \citet{PHATdust}.

\subsection{Did we find the faintest WRs in this new survey?}
Given that our previous Mosaic survey did not go deep enough to find the faintest WRs (despite our assurances at the time, see Section 4.2 in \citealt{M31WRs}), we will now re-examine this question for our current survey. \citet{MJ98} emphasize that the identification of WR candidates through narrow-band imaging is a line-flux vs.\ signal-to-noise (S/N) issue. For instance, a bright WN star with weak He\,{\sc ii} $\lambda$4686 emission will have a small magnitude difference between the {\it WN} and continuum images; thus, whether or not it is detected either by photometry or by image subtraction, becomes a S/N question. However, at the faint end, we can simplify the analysis by comparing the distribution of WR magnitudes through the on-band {\it WN} filter with the limiting magnitude of our new survey.

To determine the magnitudes of the WRs through the {\it WN} filter, we performed aperture photometry on all of the original Mosaic {\it WN} images as well as the new LMI {\it WN} images. The instrumental magnitudes (determined using a 3-pixel radius aperture) were then compared to the LGGS {\it V}-band magnitudes for all the stars categorized as being relatively uncrowded. The median differences were used to then obtain calibrated {\it WN} magnitudes tied to the {\it V}-band for all the known and newly found WRs. Results from individual images were averaged, star-by-star.

We then determined our ``limiting magnitude" for both our old and new surveys. We found that on the Mosaic frames, the faintest WRs that were identified on a single dither exposure had integrated counts of about 150 ADUs (420 e-). Combining images from all three dithers resulted in a S/N$\sim$35, ignoring the minor noise contributions from read-noise and sky. Thus, we defined the limiting magnitude based on photon counts corresponding to a S/N of 35, taking into account the detector gains and the number of images averaged (3 for Mosaic, 1 or 2 for LMI). We find that the median limiting magnitude for our Mosaic {\it WN} exposures was 21.5 mag, while our LMI {\it WN} exposures was 23.75, a full 2.25 magnitudes deeper. This is partially explained by the the longer exposure times for the LMI data compared to Mosaic (2$\times$1200s vs.\ 3$\times$300s), partially due to the generally better seeing (0\farcs9 for LMI, vs.\ a median of 1\farcs3 for the Mosaic data), and partially to the better throughput of the LDT+LMI optical system vs.\ that of the KPNO Mayall with the Mosaic camera.

\begin{figure}
\center
\includegraphics[width=0.48\textwidth]{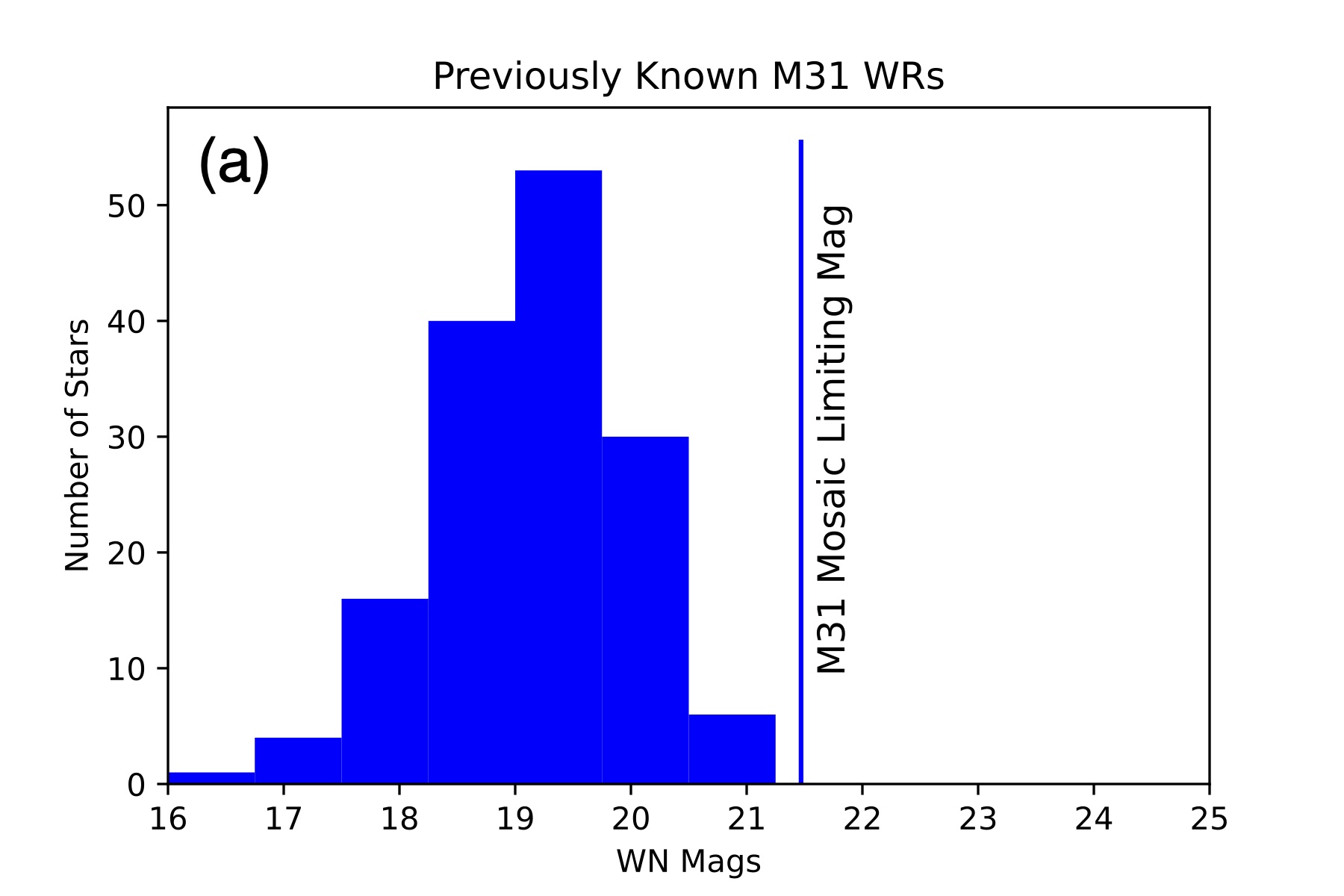}
\includegraphics[width=0.48\textwidth]{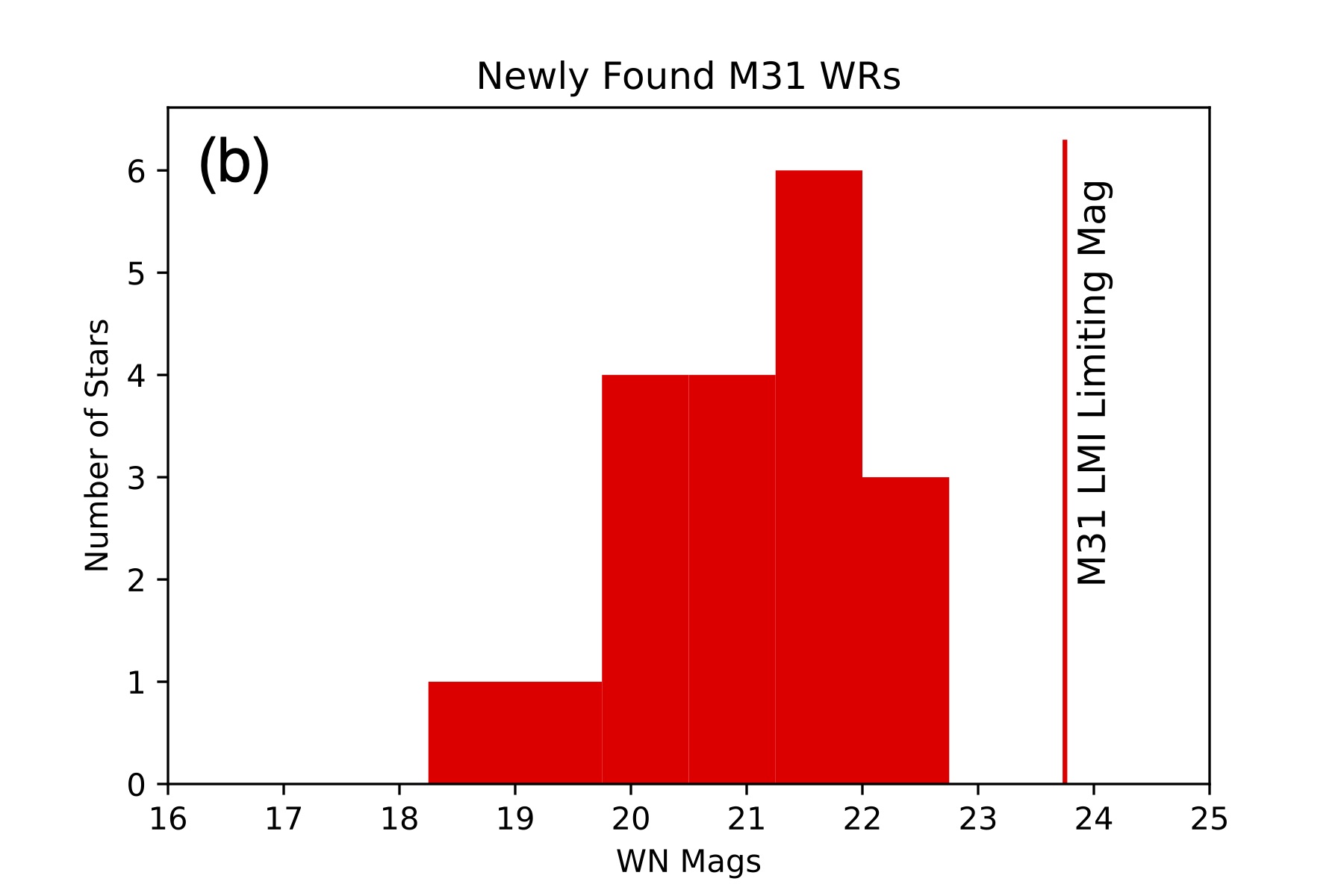}
\includegraphics[width=0.5\textwidth]{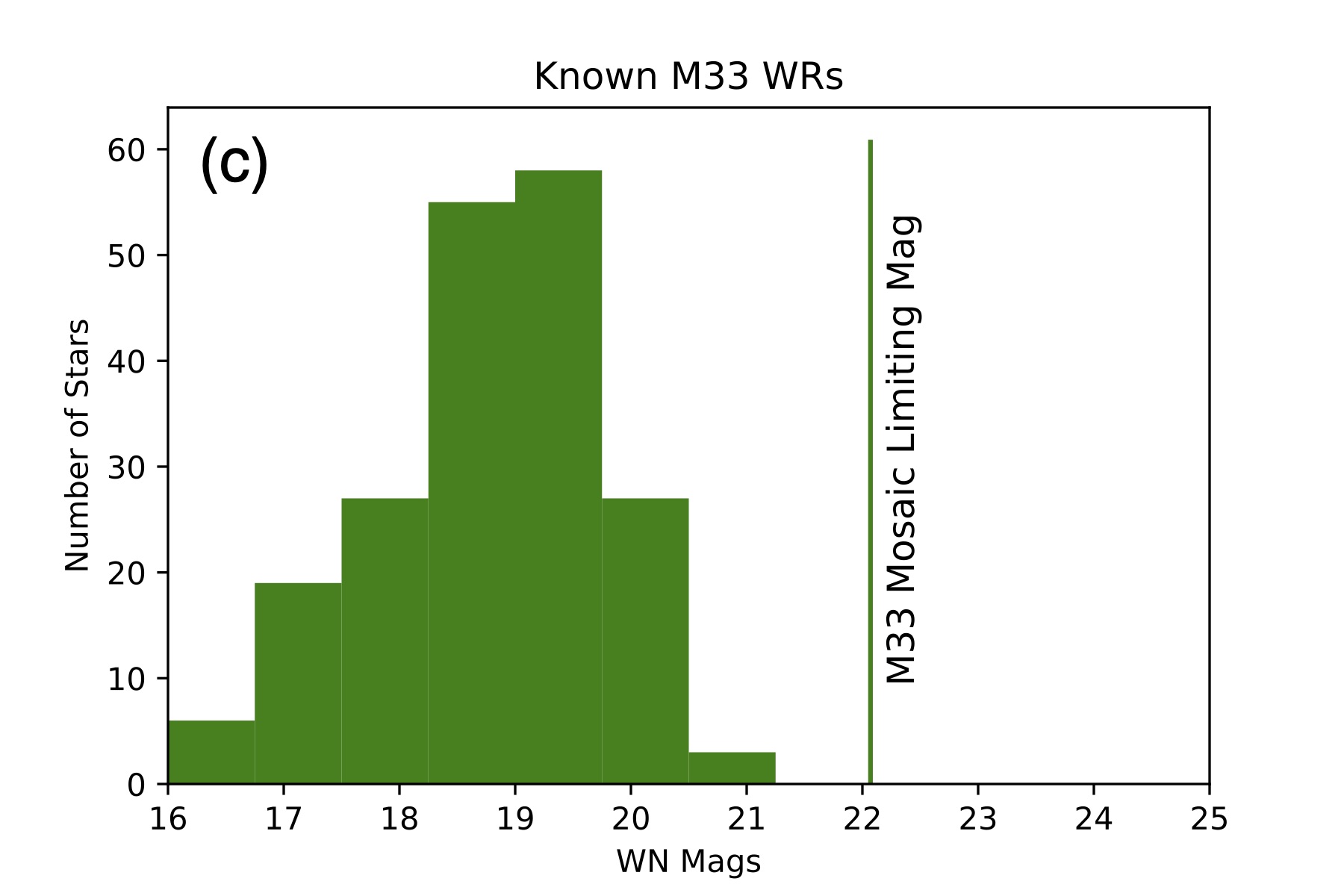}
\caption{\label{fig:histme} Distribution of {\it WN}-filter magnitudes of WRs in M31 and M33. The first (a) histogram shows the magnitude distribution of the previously known WRs in M31, mostly found in the \citet{M31WRs} Mosaic camera survey. The second (b) histogram shows the distribution of the {\it WN} magnitudes of the newly found WRs in this survey; we have also included the magnitude we measured for the star found by \citet{SharaStar}. The third (c) histogram shows the distribution of the known WRs in M33, mostly found by \citet{M33WRs} but also including the handful of WRs found subsequently (e.g., \citealt{BinaryFrequency} and \citealt{Massey2016}). In all three histograms, the vertical line shows our estimated limiting magnitude based on a S/N of 35. The data suggests that the Mosaic M31 survey was incomplete due to sensitivity, but that our new LMI frames go deep enough. In contrast, there are unlikely to be many new WRs to be found in M33.}
\end{figure}

The results are shown in panels (a) and (b) of Figure~\ref{fig:histme}. It is clear from comparison of the two histograms that the newly found WRs are, for the most part, fainter than the previously known M31 WRs. Indeed, the median {\it WN} magnitude of the 19 newly confirmed WRs is 21.0, compared to the median {\it WN} magnitude of the 150 previously known WRs of 19.2. The faintest WRs in the previously known sample have a {\it WN} magnitude of 21.6, similar to the 21.5 mag that we estimated is the limiting magnitude of the Mosaic survey. The median {\it WN} magnitude of the five faintest Mosaic WRs is 21.3. In contrast, the faintest {\it WN} magnitude in the newly discovered sample is 23.0, and the median of the five faintest is 22.8. {\it These are significantly brighter than our estimated 23.75 limiting magnitude for the LMI data.}

The large gap between the magnitudes of the faintest WRs found on our LMI images, and the limiting magnitude of those frames, suggests that if we had gone even deeper with the new data, we would not have discovered additional WRs in these fields. Rather, at least in these fields, we are sensitive enough to have discovered {\it all} of the WRs. Of course, with there may be other fields in M31 where the average reddening is greater. With only 19 newly found WRs, we may also be suffering the effects of small-number statistics. We plan on extending our survey to additional fields in M31 using LMI in the next observing season in order to further test this. For now, however, our analysis suggests that our newly found WRs allow us to make a meaningful estimates of the total number of WRs still to be found in M31, as we do in the next Section. Furthermore, we note that if our Mosaic survey needed to go about 1.5 magnitudes deeper (a factor of 4 in flux) to have reached the same limiting magnitude.

\subsection{Estimating the total number of WRs in M31}
As part of this updated survey, we only covered $\sim8$\% of the total optical disk of M31. If we assume that we will continue to find WRs at the same rate across the rest of the galaxy, that means we're missing almost 250 more WRs! However, as is shown in Figure~\ref{fig:fields} and discussed more extensively in \citet{M31WRs}, WRs aren't evenly distributed across M31. Instead, they're generally confined to their birthplaces (dense, OB forming regions) due to their high masses and short lifetimes. While we plan to continue our M31 survey in the upcoming observing season and answer this question conclusively, we can still place a few initial constraints on the total number of WRs we expect to find. 

Overall, we found approximately the same percentage ($52\pm5$\%) increase in the total number of WRs on each of the three fields we surveyed (57\%, 45\%, and 55\%). If we assume this percentage increase will remain constant for the remaining portions of M31, we can crudely estimate that there are $\sim 80 \pm 8$ WRs missing from our previous Mosaic survey. Since we have now discovered 19 new WRs as part of this survey, there are $\sim 60 \pm 8$ more WRs left to be discovered M31. These numbers clearly only serve as {\it very} broad estimates, and we look forward to constraining this number further with new observations. As mentioned previously, we purposely chose to focus this survey on areas where the Mosaic images were lacking either due to chip gaps or poor sky conditions. Therefore, this estimation for the number of missing WRs is likely a slight overestimate, rather than an underestimate. 

We do note that while the total number of WRs increased as part of our recent observing campaign, the overall ratio of WC to WN-type WRs stayed relatively constant. As summarized in \citet{NeugentGalaxies}, the observed ratio of WC to WN-type WRs as a function of metallicity is an exacting test of stellar evolutionary models and recently the observations have been in good agreement with theoretical predictions. Before discovering these new WRs, the WC/WN ratio in M31 was $0.67\pm0.11$. With the addition of 19 new WRs, the WC/WN ratio is still within the previously quoted error and is now $0.63\pm0.10$. So, while the total number of WRs has increased, the WC/WN ratio remains essentially unchanged. Additionally, although our Mosaic survey of M31 may have missed some of the fainter, more reddened stars, it was a major step foward in our knowledge of the WR content of our nearest spiral neighbor. Prior to \citet{M31WRs}, only 47 WRs were known, mostly of WC type. Their Mosaic survey added another 107, and brought the observed ratio of WC/WN stars to a value that could be compared to the evolutionary models.

\subsection{What does this mean for M33?}
As mentioned above, our discovery does raise questions about the completeness of the similar Mosaic WR survey done in M33 \citep{M33WRs}. Panel (c) in Figure~{\ref{fig:histme} shows the distribution of the {\it WN} magnitudes of the M33 WRs. The limiting magnitude of the M33 Mosaic survey is significantly fainter than for the M31 Mosaic survey, 22.1 vs.\ 21.5. The reason is easy to understand: the M33 data were taken under better conditions, with a median seeing of 1\farcs0 rather than 1\farcs2. The faintest WR star has a {\it WN} magnitude of 20.8; the median value of the faintest five WRs is 20.6~mag. These are all significantly brighter than the 22.1~mag limiting magnitude. Based on this, we do not expect there to be a significant population of missing WRs in M33.  That said, we are in the process of completing a spectroscopic followup of M33 WR candidates found as part of a new survey with the LDT and LMI. The majority of additional candidates were located in the Mosaic gaps, but a few fainter candidates exist. We will report fully on this when the study is complete.

\section{Summary and Conclusions}
After several indications emerged that our previous WR survey in M31 discussed in \citet{M31WRs} may have been incomplete (e.g., \citealt{SharaStar}, \citealt{RSGWR}), we began a pilot survey with the goal of both finding new WRs and better understanding the completeness limits of our previous survey. The major results of this pilot survey are summarized as follows:
\begin{enumerate}
\item We spectroscopically confirmed 19 new WRs in M31.
\item Three of these new WRs were in the gaps of our previous survey. One of them was in the previous survey but too crowded to be readily identified. The remaining 15 new WRs were missed on the previous survey due to their faint magnitudes.
\item These 15 faint WRs are not intrinsically faint, but rather have slightly increased reddening compared to the WRs in M31 that were discovered as part of our previous survey.
\item We estimate that there are around 80 WRs missing from our previous survey, 19 of which have been discovered as part of this work. Thus, we estimate that there are around 60 WRs left to be found in M31.
\item After calculating the limiting magnitude of our new survey, and comparing it to the magnitudes of the WRs we discovered, we conclude that this new survey is not missing another even fainter population of WRs in M31.
\item We additionally conclude that our previous survey of WRs in M33 discussed in \citet{M33WRs} did not suffer from the same completeness issues. While there are likely a few fainter WRs left to be discovered in M33, the issue is not as severe as with M31.
\end{enumerate}

We look forward to continuing this survey in the upcoming observing seasons. Besides finding new WRs, we will also obtain updated $B$,$V$,$R$ photometry to better estimate the effect of reddening on our completeness limits. Our updated sample will better constrain the RSG/WR ratio discussed in \citet{RSGWR}, allow for more accurate comparisons with evolutionary models, and ultimately help us better understand the evolution and end states of massive stars.

\acknowledgements
We thank the anonymous referee for suggestions that improved the paper, as well as Ben Weiner for his assistance when designing the Binospec masks. P.\ M.\ is also grateful to his Lowell Observatory colleague Larry Wasserman for invaluable help in the astrometric refinements of the WR coordinates that enabled us to observe these stars spectroscopically with Binospec. 

Additionally, we acknowledge that the work presented was done on the traditional territories and ancestral homelands of the Cheyenne, Arapaho, Ute and many other Native American nations and that Lowell Observatory sits at the base of mountains sacred to tribes throughout the region. We honor their past, present, and future generations, who have lived here for millennia and will forever call this place home.

Support for this work was provided by NASA through the NASA Hubble Fellowship grant \#HST-HF2-51516 awarded by the Space Telescope Science Institute, which is operated by the Association of Universities for Research in Astronomy, Inc., for NASA, under contract NAS5-26555. We also thank the Mt.\ Cuba Astronomical Foundation for their generous support, which enabled purchase of the customized interference filters used for this survey.

Observations reported here were obtained at the MMT Observatory, a
joint facility of the University of the Smithsonian Institution and
the University of Arizona.  These results also made use of the
Lowell Discovery Telescope (LDT) at Lowell Observatory.  Lowell is
a private, non-profit institution dedicated to astrophysical research
and public appreciation of astronomy and operates the LDT in
partership with Boston University, the University of Maryland, the
University of Toledo, Northern Arizona University, and Yale University.
The Large Monolithic Imager was built by Lowell Observatory using
funds provided by the National Science Foundation (AST-1005313).
This research has made use of the SIMBAD database, operated at CDS,
Strasbourg, France. This research has made use of the VizieR catalogue
access tool, CDS, Strasbourg, France (DOI: 10.26093/cds/vizier).
The original description of the VizieR service was published in
A\&AS 143, 23. This work has also made use of data from the European
Space Agency (ESA) mission {\it Gaia}
(\url{https://www.cosmos.esa.int/gaia}), processed by the {\it Gaia}
Data Processing and Analysis Consortium (DPAC,
\url{https://www.cosmos.esa.int/web/gaia/dpac/consortium}). Funding
for the DPAC has been provided by national institutions, in particular
the institutions participating in the {\it Gaia} Multilateral
Agreement.

This work made use of the following facilities and software:
\facility{LDT, MMT, {\it Gaia}}

\software{IRAF (distributed by the National Optical Astronomy Observatory, which is operated by the Association of Universities for Research in Astronomy under a cooperative agreement with the National Science Foundation), Python 3.7.4}

\appendix
The following are notes on the individual spectroscopically observed WR candidates.

\textbf{J004019.66+404232.5.} This is an early-type WN star with He\,{\sc ii}~$\lambda4686$ and N\,{\sc v}~$\lambda\lambda4603,19$. N\,{\sc iv}~$\lambda4058$ is also present, but weaker than the N\,{\sc v} doublet,  making this a WN3. There are strong nebular lines, primarily the [OIII]~$\lambda\lambda4959,5007$ doublet plus Balmer emission. 

\textbf{J004031.21+404128.1.} This is hydrogen-rich WN7 star, with N\,{\sc iii}~$\lambda4634,42$ emission about half as strong as He\,{\sc ii}~$\lambda4686$. N\,{\sc iv}~$\lambda4058$ is strongly present but weaker than N\,{\sc iv}. The even-N Pickering lines of He\,{\sc ii}, which are coincident with the Balmer lines, are much stronger than the odd-n lines, revealing the presence of hydrogen. 

\textbf{X004032.57+403901.3.} This is a WC6 star, with C\,{\sc iii/iv}~$\lambda4650$ and C\,{\sc iv}~$\lambda\lambda5801,12$ as the strongest lines. C\,{\sc iii}~$\lambda5696$ is readily visible, weaker than C\,{\sc iv}, and O\,{\sc v}~$\lambda5592$ is weaker still.  Balmer absorption from H$\delta$ short-wards is visible, as is He\,{\sc i}~$\lambda4471$, suggesting a OB-type companion, either physical or line-of-sight.  The fact that the C\,{\sc iv}~$\lambda4650$ line has an EW of -40\AA\ (rather than $<-100$\AA) is consistent with significant dilution by such a companion. (See the WC6 star X004054.05+403708.3 described below, with an EW of -900\AA.)

\textbf{J004039.59+404449.5.} Our spectrum nicely identifies this as a WN3 star, with He\,{\sc ii}~$\lambda4686$, N\,{\sc v}~$\lambda\lambda4603,19$, and N\,{\sc iv} $\lambda4058$. The N\,{\sc v}~$\lambda4946$ line is also present. 

\textbf{J004042.44+404505.3.} In this star N\,{\sc iv}~$\lambda4058$ and N\,{\sc v}~$\lambda\lambda4603,19$ are the dominant nitrogen ions, with N\,{\sc iv} of slightly greater intensity. N\,{\sc iii}~$\lambda\lambda4634,42$ is not present. Together, these comparisons lead to a WN4.5 classification. C\,{\sc iv}~$\lambda\lambda5801,12$ is also present, although not at the intensity that would cause us to designate this a WN/C star.  He\,{\sc ii} lines at 4100\AA, 4542\AA, 5411\AA, and 6560\AA\ are all also visible with a fairly smooth progression and there is no evidence of hydrogen.

\textbf{X004045.57+404526.4.} Our spectrum of this candidate shows only an inverse nebular spectrum, suggesting over-subtraction of an H\,{\sc ii} region. No continuum or WR features are evident in our spectra.

\textbf{X004054.05+403708.3.} The star has little continuum, but the strong C\,{\sc iii/iv}~$\lambda4650$ and C\,{\sc iv}~$\lambda\lambda5801,12$ lines stand out.  C\,{\sc iii}~$\lambda5696$ and O\,{\sc v}~$\lambda5592$ are of similar intensity, but with C\,{\sc iii} a bit stronger, and we thus classify this star as a WC6.

\textbf{X004318.88+414711.1.} This is an H\,{\sc ii} region with nebular He\,{\sc ii}~$\lambda4686$ emission.

\textbf{X004320.88+414107.0.} This is a late-star, with TiO bands present at 5167\AA, 6158\AA, and 7054\AA, probably a K7-M0. The star lacks Gaia DR3 proper motions or parallaxes, but the radial velocities of the Ca\,{\sc ii} triplet in our spectrum supports membership.  The star does not appear in Table~2 of \citet{Massey23} presumably as it is too faint.

\textbf{J004326.06+414260.0.} The star shows N\,{\sc v}~$\lambda\lambda4603,19$ with comparable N\,{\sc iv}~$\lambda4058$, but no N\,{\sc iii}~$\lambda\lambda4634,42$. Thus we classify it as WN4. There are strong He\,{\sc ii} lines, but a smooth progression from even-n to odd-n giving no evidence of hydrogen. A strong nebular spectrum is also present.  C\,{\sc iv}~$\lambda\lambda5801,12$ is also present, but again not at a level that we would call this a WN/C. 

\textbf{X004332.04+414817.2.} This is an H\,{\sc ii} region with nebular 4686\AA\ emission.

\textbf{X004338.95+414327.0.} The spectrum is dominated by nebular emission, but there is clearly a broad component at 4686\AA. The star is of WN-type but we cannot classify it more exactly. 

\textbf{X004341.15+414413.6.} Our two spectra of this star are compromised by a strong over-subtraction near the 4650-4686\AA\ complex, some sort of reduction issue. Nevertheless, we can classify this as a late-type WN, with N\,{\sc iii}~$\lambda\lambda4634,42$ equal in intensity to He\,{\sc ii}~$\lambda4686$. The star is clearly of WN-type rather than an Of star, as He\,{\sc i}~$\lambda5876$. H$\alpha$ is also present in emission. C\,{\sc iv}~$\lambda\lambda5801,12$ is as strong as He\,{\sc ii}~$\lambda4686$, and so we call this an WN8/C star.

\textbf{X004353.03+412141.0.} This is a late-type WC, as evidenced both by the fact that C\,{\sc iii}~$\lambda5696$ is comparable in strength to C\,{\sc iv}~$\lambda\lambda5801,12$, and the fact that the lines are relatively skinny for a WC. We classify this as a WC7.  Both the line fluxes and the continuum are weak; with a less powerful instrument or shorter exposure times this star would not have been detected as a WR. A strong nebular spectrum is also present.

\textbf{X004359.44+414823.9.} We classify this as a WN6, as N\,{\sc iii}~$\lambda\lambda4634,42$ is about the same intensity as N\,{\sc iv}~$\lambda4058$, with N\,{\sc v}~$\lambda\lambda4603,19$ weakly present. He\,{\sc ii} emission is weakly present. 

\textbf{J004404.10+411710.5.} This is an H\,{\sc ii} region but without He\,{\sc ii}~$\lambda4686$ emission. It is not clear why this star was selected as a WR candidate.

\textbf{X004406.41+412020.8.} N\,{\sc iv}~$\lambda4058$ and N\,{\sc v}~$\lambda\lambda4603,19$ are the dominant N ions in this WN star, with N\,{\sc iii}~$\lambda\lambda4634,42$ very weakly present. N\,{\sc iv} is slightly stronger than N\,{\sc iii}, We therefore classify this as a WN4.5. There is no odd-n, even-n discrepancies among the Pickering lines, and so the star shows no obvious evidence of hydrogen. C\,{\sc iv}~$\lambda\lambda5801,12$ is weakly present.

\textbf{J004408.13+412100.6.} The lines are sharp and strong in this late-type WN star.  N\,{\sc iv}~$\lambda4058$ is the dominate N ion, but in similar intensity to N\,{\sc iii}~$\lambda\lambda4634,42$, with N\,{\sc v}~$\lambda \lambda4603,19$ quite visible but much weaker, leading to a WN4.5 classification. Strong nebular lines are present.

\textbf{J004413.56+412004.7.} This is another WN4.5 star, with N\,{\sc iv}~$\lambda4058$ marginally stronger than N\,{\sc v}~$\lambda\lambda4603,19$ and N\,{\sc iii}~$\lambda\lambda4634,42$ absent. The He\,{\sc ii} emission sequence does not suggest the presence of hydrogen.  

\textbf{X004414.71+414033.3.} This is an interesting but odd discovery.  The object is not a WR star, but narrow, nebular-like emission is present at He\,{\sc ii}~$\lambda4686$ as well as H$\alpha$, H$\beta$, H$\gamma$, and H$\delta$. However, the expected forbidden lines of O[{\sc iii}]~$\lambda\lambda4959,5007$ are not present. TiO band absorption at 5167\AA, 6158\AA, and 7054\AA\ and a red continuum suggests an underlying M0-ish star present.  We believe this is a symbiotic star. There is also strong emission present at a rest wavelength of 5583$\pm$1\AA, which we were unable to identify.

\textbf{X004418.10+411850.8.} There are only two WR features present in our spectrum, broad C\,{\sc iii/iv}~$\lambda4650$ and O\,{\sc vi}~$\lambda\lambda3811,34$.  The latter is even stronger than the C\,{\sc iii/iv} making this star a WO-type. No WR lines are visible in the yellow in our spectrum, and so we cannot give a more exact subtype. Strong nebular emission is seen.

\textbf{X004421.30+411807.2.} The star appears to be of ``Of" type, with N\,{\sc iii}~$\lambda\lambda4634,42$ and He\,{\sc ii}~$\lambda4686$. The EW of the latter is -4\AA. The pipeline reduction has over-subtracted nebular emission, but He\,{\sc i}~$\lambda4387$ and $\lambda4471$ absorption are present, as well as He\,{\sc ii}~$\lambda4200$ and $\lambda4542$ absorption, in addition to Balmer absorption. We classify the star as O6~If.

\textbf{X004423.98+412255.6.} This is a cool star, with TiO bands at 5167\AA, 5847\AA, 6158\AA, 6658\AA, and 7054\AA. The presence of the 5847\AA\ and 6658\AA\ suggests a spectral subtype later than M0, probably more like an M2.  There are no Gaia data for this star, but an examination of the the Ca\,{\sc ii} triplet shows a radial velocity consistent with M31 membership. The star is not included in Table 2 of \citet{Massey23}, presumably because it is too faint.

\textbf{X004425.09+412046.4.} This is a classical WC6 emission spectrum on a strong continuum with O-type absorption lines at shorter wavelengths. We tentatively classify the companion as O8.

\textbf{X004426.99+411928.0.} There is a relatively early-type absorption present, but with no sign of He\,{\sc i} or He\,{\sc ii}. There is no He\,{\sc ii}~$\lambda4686$. This star was included as a WR candidate likely because it was too bright for image subtraction to work correctly (see discussion in Section 2.1.2).

\textbf{X004431.39+412114.0.} N\,{\sc v}~$\lambda\lambda4603,19$ is the dominate nitrogen line in this star, with very weak N\,{\sc iii}~$\lambda\lambda4634,42$ present.  No N\,{\sc iv}~$\lambda4058$ is visible but the spectrum is very noisy in the far blue.  We classify this star as a WN3. A strong nebular spectrum is also present.

\textbf{X004432.06+411940.5.} This is an H\,{\sc ii} region with He\,{\sc ii}~$\lambda4686$ emission. There is also an absorption line spectrum present, mainly evidenced in the upper Balmer region. 

\textbf{J004433.91+412501.2.} N\,{\sc iv}~$\lambda4058$, N\,{\sc v}~$\lambda\lambda4603,19$, and N\,{\sc iii}~$\lambda\lambda4634,42$ are all present in our spectrum of this star, with N\,{\sc iv} the strongest, and N\,{\sc v} the weakest. We call this a WN6 type. C\,{\sc iv}~$\lambda\lambda5801,12$ is also present at modest intensity. Strong nebular lines are present.

\textbf{X004438.04+412518.7.} The spectrum of this star is very similar to that of J004433.91+412501.2, except no C\,{\sc iv} is evident.  We classify this as a WN6.

\textbf{X004440.49+412052.1.} This is another H\,{\sc ii} region with He\,{\sc ii}~$\lambda$4686 emission.

\newpage
\bibliographystyle{apj}
\bibliography{masterbib}

\end{document}